%% file: main_paper_2018.tex
\documentclass{llncs}
\usepackage{makeidx}  
\usepackage[english]{babel}
\usepackage[utf8]{inputenc}

\usepackage{multicol}
\usepackage{boxedminipage}

\usepackage{graphicx}
\usepackage{fancybox}
\usepackage{pdfpages}

\usepackage{amsmath}

\usepackage{diagrams}
\usepackage{alltt}
\usepackage{wrapfig}

\usepackage{array}

\usepackage{boxedminipage}
\usepackage{multicol}
\usepackage[pdftex]{thumbpdf}

\usepackage[pdfusetitle]{hyperref}
\usepackage{cleveref}
\usepackage{tikz}

\usepackage{xcolor}
\usepackage{todonotes}
\usepackage{menukeys}
\usepackage{amssymb}

\usepackage{zedbis}
\usepackage{b2latex}

\usepackage{color}

\presetkeys{todonotes}{color=green!20}{}

\newcommand{\pow}{\mathbb P\hbox{}}

\definecolor{keycolor}{rgb}{0,0,0.8}     
\newcommand{\bkey}[1]{\textcolor{keycolor}{#1}}
\definecolor{cmtcolor}{rgb}{0,0.4,0}     
\newcommand{\bcmt}[1]{\textcolor{cmtcolor}{\small{#1}}}

\begin{document}
%
%
\pagestyle{headings}  

%
%
%
\title{Building Correct SDN-Based Components from a Global Formal Model}

\titlerunning{Global SDN Formal Model}  
%
\author{J. Christian Attiogb{\'e} \\[0.7ex]  
}
\institute{LS2N  - UMR CNRS 6004 - University of Nantes\\
\email{Christian.Attiogbe@univ-nantes.fr}\\
}


\maketitle     

\begin{abstract}
  Software Defined Networking (SDN) brings flexibility in the construction and managment of distributed applications by reducing the constraints imposed by physical networks and by moving the control of networks closer to the applications. However mastering SDN still poses numerous challenges among which the design of correct SDN components (more specifically controller and switches). In this work we use a formal stepwise approach to model and reason on SDN.
  Although formal approaches have already been used in this area, this contribution is the first state-based approach; it is based on the Event-B formal method, and it enables a correct-by-construction of SDN components. 
We provide the steps to build, using several refinements, a global formal model of a SDN system;
  correct SDN components are then systematically built from the global formal model satisfying the desired properties. Event-B is used to experiment the approach. 
\end{abstract}

\section{Introduction}
\label{section:introduction}
\input{introduction}

\section{Overview of SDN: Concepts and Architecture}
 \label{section:sdnOverview}
\input{sdnOverview.tex}

\section{Stepwise Refinement-based Modelling of SDN}
\label{section:modelling}
\input{modelling.tex}
\section{Deriving Correct Controller and Switch Components}
\label{section:controller}
\input{controller.tex}

\section{Experimentations and Assessment}
\label{section:experimentation}
\input{experimentation.tex}

\section{Conclusion}
 \label{section:conclusion}
\input{conclusion.tex}



\input{main_paper_2018.bbl}
\end{document}

%% file: introduction.tex
%
An essential constituent of distributed applications is the physical network behind them. Distributed applications very often, build on existing middlewares which embody services provided by  the network level. Thus the reliability of distributed applications depends not only on their own development but also on the reliability of the network.
Due to the involvment of many physical devices, the network level has been for many years a source of severe complexities and constraints leading very often to the adoption of rigid solutions in the deployment of  applications. 

Fighting the lack of flexibility of physical networks has resulted in the Software-Defined Networking (SDN) initiative \cite{Kreuz2015-Survey,OpenNetworkFoundation-OF2013,OpenNetworkFoundation-OF2012WPaper}.
Software-Defined Networking now provides the opportunity to go deeper in modelling and reasonning on networks, since it enables to define and manage more easily the networks at software level. Indeed a Software-Defined Network  is made of a controller and switches which are abstractly defined before being implemented at software level.
In this context an user application does not consider a physical network or a specific middleware but it is rather  built on top of a virtual or open network.

Even if Software-Defined networking  makes it possible to control an entire network with software, through programs that tailor network behavior to suit specific applications and environments, programmers still have many difficulties to write correct SDN programs. This is due to the unpredictability of the SDN as a distributed asynchronous system, and the lack of correctly developped SDN frameworks or formally verified SDN frameworks.
Many works have been undertaken around SDN; they address different aspects: building simulators and analysers for SDN, building SDN controllers, verifying the controller component of an Software-Defined Network, etc.

However SDN deployment is still at its beginning and programmers or administrators still need trustworthy materials and frameworks.
Such materials may come from rigorously founded models and related reasonning and engineering tools.
Besides, considering the keen and the demands for the deployment of SDN as a flexible infrastructure for specific applications, clouds applications, IoT, etc, which all require security, it is of tremendous importance to have at the disposal of developpers trustworthy development, analysis and simulation  frameworks. Formal models taking into account several of these aspects are then necessary. That is the challenges that motivates of our work.

The main contributions of this paper are:
\textit{i)} capturing the SDN as a discrete-event system to foster its modelling with an event-based approach;
\textit{ii)} a state-based core model for rigorous analysis, development and simulation frameworks dedicated to SDN applications. It is a global Event-B \cite{EventB_Abrial2010} formal model, designed as the basis of the stepwise construction of the various components of a SDN; 
\textit{iii)} the systematic derivation of correct components (SDN controller and switches) from the global model which is previously proved to have some required properties.

The article is organised as follows. Section \ref{section:sdnOverview} gives an overview of the Software-Defined Networking,  related works and main issues.
In Section \ref{section:modelling} we introduce the main concepts for modelling SDN, an overview of Event-B method and then our approach to build the global abstract model by stepwise refinements.
Section~\ref{section:controller} shows how one can derive the construction of a correct SDN controller from the global formal model.
Section~\ref{section:experimentation} gives the first experimental results related to simulation and verification of global safety/liveness properties.
We conclude in Section~\ref{section:conclusion} and stress some challenges for future work.

%% file: sdnOverview.tex
\begin{figure}
  \begin{center}
\includegraphics[width=0.6\linewidth]{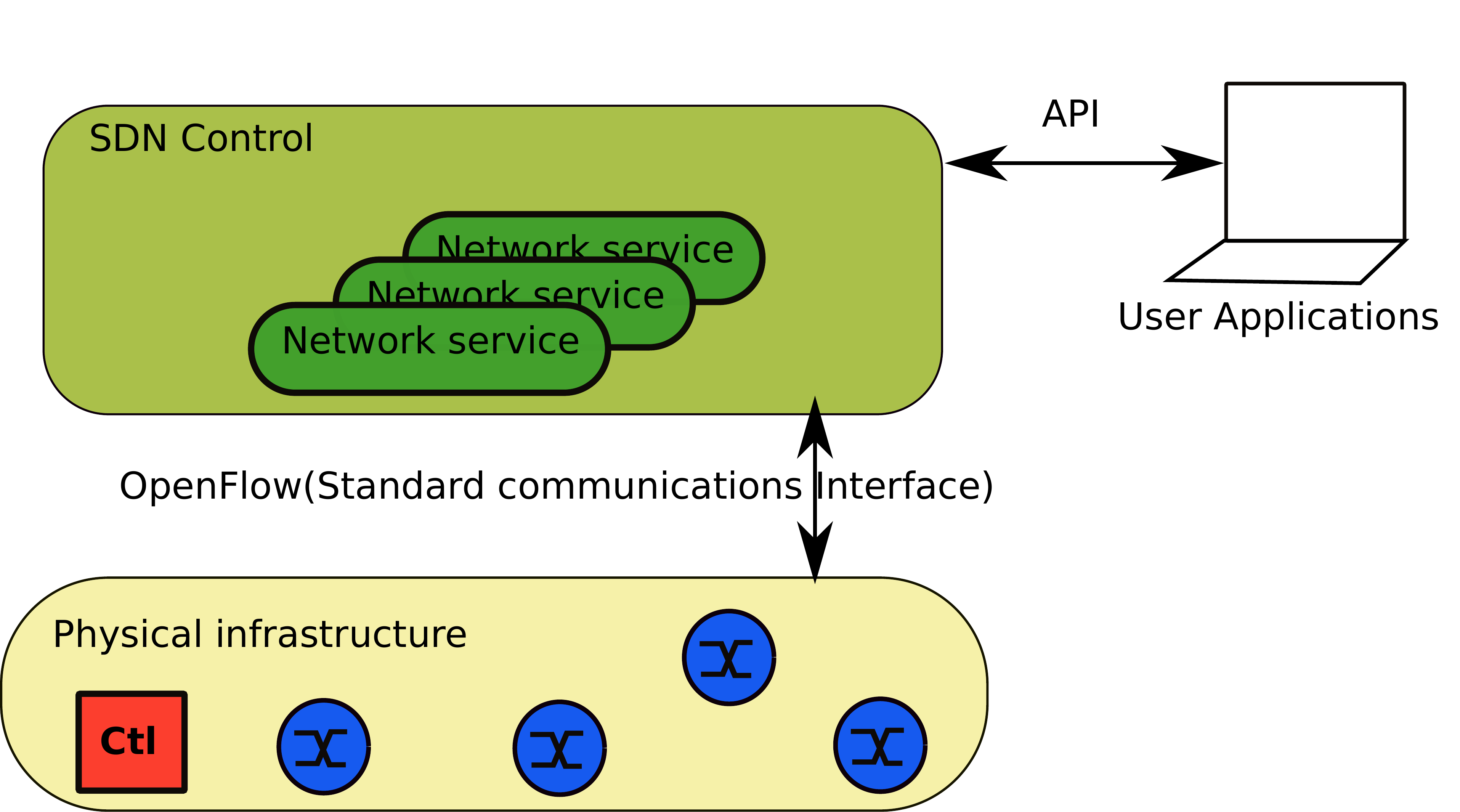}
  \caption{The layered architecture of a SDN}
  \label{figure:archiSDN}
\end{center}
\end{figure}

%
%
The SDN architecture consists of three layers: User-application, Control and Data forwarding. Control and Data are the most relevant ones when studying the SDN.
Figure \ref{figure:archiSDN} depicts how the SDN is viewed from the user side as a single global switch which denotes an abstraction of an entire network. User applications can directly exploit an abstraction of the network. Network services are solicited directly from host machines linked to a physical device: a \textit{switch} assumed to be under the SDN control.

In Software-Defined Networking there is a physical separation between the \textit{control-plane} (the management of the network and definition of rules to forward data packets) and  the \textit{data-plane} (how packets are forwarded according to control-plane) \cite{OpenNetworkFoundation-OF2013,Kreuz2015-Survey}. 
Indeed the network control (the high level or control-plane) is now separated from the packet forwarding (the low level or data-plane) activity and, physical devices inside the low level may be designed more easily; the network control is independent from device providers, the control is brought closer to software controllers and administrators.
Traditional network services such as routing, access control, traffic engineering, security, etc can be implemented via various API provided by the SDN, instead of being vendor-dependent.
The control and data levels are linked by an open communication interface.
OpenFlow \cite{OpenNetworkFoundation-OF2012,OpenNetworkFoundation-OF2012WPaper} is representative of such communication interfaces.
OpenFlow is a standard communication interface, that moves the control of the network from the physical \textit{switches}  to logically centralized control software.

SDN has been used in variety of implementations, for example \cite{OF_wireless_2011} is dedicated to the implementation of wireless networks, while in \cite{FlowCheckerAl-Shaer:2010:FCA:1866898.1866905} the authors describe a tool, FlowChecker, which identifies any intra-switch misconfiguration within a FlowTable of a switch.
RouteFlow \cite{RouteFlow2014} is a controller which implements virtualized IP routing over OpenFlow infrastructure.

\subsection{Concepts and Components}
\begin{figure}
  \begin{center}
\includegraphics[width=0.6\linewidth]{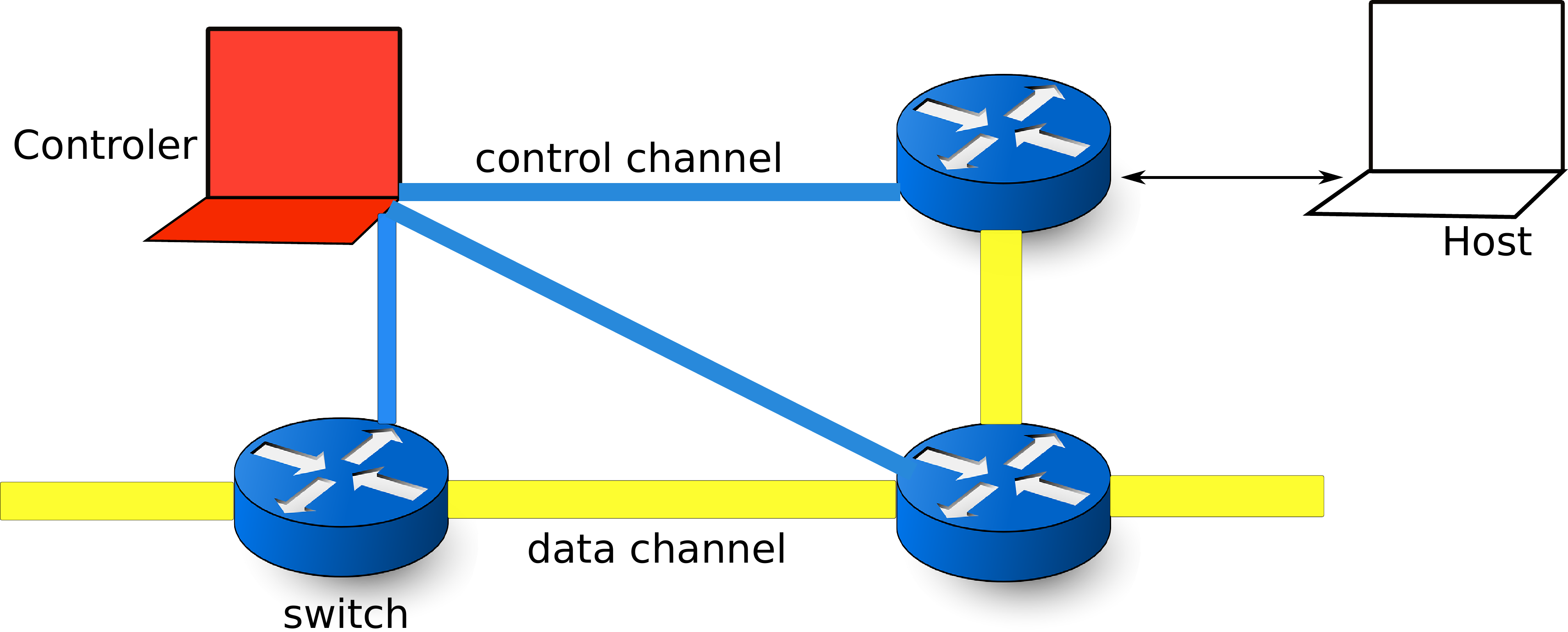}
  \caption{A detailed architecture}
  \label{figure:detailedArchiSDN}
\end{center}
\end{figure}

We distinguish in Figure \ref{figure:detailedArchiSDN}, the main components of an SDN.
Switches and controller are network devices that interact using packets and messages on data channel and message channel.

\paragraph{Switch.}
A switch is a device responsible of forwarding packets, to perform a hop-by-hop transfer of information through the network. A switch is configurable  by the controller with which  it interacts with the controller via a secure message channel. A switch interacts with other switches via a data channel.
\paragraph{Controller.} A controller is a  device responsible of controlling a whole network (a local or medium area network). It is used by the network administrator to dynamically configure, in an evolutive way, the switches with adequate forwarding information; it maintains the connectivity of the switches, etc.
The controller  initiates the switches behaviour, maintains them and instructs the switches with respect to specific actions. Packets not treated by the switches are sent to the controller via messages emitted on the  secure channel. The controller does not use the data channel.
\vspace{-0.2cm}
\paragraph{Packet.}
A packet contains information related to various protocols (Ethernet, IP, etc). A packet has a header related to Data, Networkand layers and a body. Inside the header we have for instance: the destination and source addresses for each layer, the type of protocol,  ...
\paragraph{Message.}
A message contains a control or management information addressed by the controller to a switch. The control information is for instance:  which packet to drop, the indication of a port on which the switch may forward a data packet. 
A switch can also emit a message to a controller. In this case either the message contains a response to a control order (for instance the controller asked for the status of a switch) or a packet for which the switch does  not have an entry in its table for forwarding the packet to its destination.
\vspace{-0.2cm}
\paragraph{Flow Table.} A flow table is a part of a switch. It describes the switch elementary behaviour. A flow table is made of several entries sent by the controller. Each entry has a header information and a body. The header may contain a message priority set by the controller. The body of the message can be a data packet, or a rule to process the incoming packets.
\vspace{-0.2cm}
\paragraph{Interaction between switches and controller.}
A properly configured switch  has routes to forward received packets coming from other network services. If the switch lacks of forwarding information, it sends the received packets to the controller. The controller is linked to the available switches and manages them directly with orders sent via messages on a \textit{secure reliable channel}.  These messages are used to configure and maintain the switches, defining for each one the rules to forward packets it receives.  
At this stage we have a simple interaction between application level, and the provided high level network services. But this interaction is more complex if we look at it in details.
Consider for this purpose a detailed view of the interaction with the SDN network depicted by Figure~\ref{figure:detailedArchiSDN}.

\subsection{OpenFlow: a Standard Interface}
OpenFlow is a standard communications interface, supported by the Open Network Foundation \cite{OpenNetworkFoundation-OF2012,OpenNetworkFoundation-OF2013,OpenNetworkFoundation-OF2012WPaper}. OpenFlow has been precisely defined but not formally.

As such, OpenFlow provides a means for specifying data level or control-plane logic and also protocols.  However there is no mandatory formal specification or formal requirements; accordingly the network systems resulting from OpenFlow may be  incorrect or not satisfying safety conditions. 

The OpenFlow semantics being unformal, tool builders may assume particular behaviour and functioning for the network devices, leading to inconsistencies and incorrect behaviours; that is the case for the order in which packets are processed inside a switch.

\subsection{Issues and Related Works}
There is a keen for SDN, justifying several works both from industry and academy. Important efforts are devoted to the implementation of SDN \cite{RouteFlow2014,OF_wireless_2011,RAZA20141050}.
SDN provides flexible network systems and distributed systems development but there is no guarantee that these systems are safe or correct.
SDN as an asynchronous system undergoes the impact of time passing and non-determinism or concurrency of events. Packets may be received and distributed in any order causing for instance inconsistent interpretation in the switches when a forwarding route arrives after the related packet is sent.  
One of the main issues in SDN is the inconsistent packet forwarding during a network update which results in an update inconsistency \cite{El-Hassany:2016:SCA:2908080.2908124}.
Update consistency requires that packets are either forwarded by an old version of the forwarding  table or by the new version of the table (after an update), but not by an interleaving of the old and the new version.

These issues require efforts to build robust tools and protocols on the basis of thoroughly studied SDN models.
Several works have been devoted to various aspects of SDN among which the modelling and reasoning on the SDN controller\cite{Guha:2013:MNC:2491956.2462178}, the analysis of the SDN traffic \cite{El-Hassany:2016:SCA:2908080.2908124,Khurshid:2013:VVN:2482626.2482630}.

According to the state of the art \cite{Kreuz2015-Survey,HORVATH2015552,AKHUNZADA2016199,ALSMADI201579-security} most investigations address the implementation issue as an important challenge; some of the aspects taken into account in these works are: scalability, performance, security, simulation.
The correction of the implementations has received less attention.

In \cite{Hu:2014:FBR:2620728.2620749} the authors address the challenge of building robust firewalls for protecting OpenFlow-based networks where network states and traffic are frequently changed. They propose the FlowGuard framework for the detection and the resolution of firewall policy violation. They use an algorithmic approach.

VeriFlow is a verification tool proposed in \cite{Khurshid:2013:VVN:2482626.2482630} for checking in real-time network invariants and policy correctness in OpenFlow networks. This work is based on direct implementation of the forwarding rules and an algorithmic approach, that monitors the update events occurring on the network.

SDNRacer \cite{El-Hassany:2016:SCA:2908080.2908124} is a network analyzer which can ensure that a network is free of harmful errors such as data races or per-packet inconsistencies.
The authors provide a formal semantics enriched with a  happens-before model, to capture when events can happen concurrently. 

The work in \cite{Guha:2013:MNC:2491956.2462178} is devoted to the verification of an SDN controller;
the authors provide an operational model for OpenFlow and formalize it in the Coq proof assistant.
This model is then used to develop a verified compiler and a run-time system for a high-level network programming language.

To sum up, there is the preliminary steps towards making SDN networks more reliable; but much works remain to be done:
\begin{itemize}
\item making it easier for developers the construction and verification of controllers from various existing well-researched models,
\item enhancing machine-assisted configuration of controllers and OpenFlow-based switches,
\item promoting the reuse of correct SDN components in the deployment of new SDN (that is interoperability).
\end{itemize}

The goal of our work is to serve these purposes by contributing with a global, extensible, refinable formal model. It is the first event-based one, making it easy to derive simulators and also to prove safety and liveness properties. It is provided as a reusable formal basis for any one interested, avoiding hence to repeat the efforts through the chains of works.

Unlike in the case dedicated to implementations, we follow an approach similar to those addressing modelling and reasoning on controllers, by defining for the SDN a global formal model  from which the models of the components can be derived and then correctly implemented.


%% file: modelling.tex
%
We  use Event-B \cite{EventB_Abrial2010} and adopt a correct-by-construction approach.
\vspace{-0.2cm}
\subsection{An overview of Event-B}
\input{eventB_nutshell-1}

\subsection{Abstractions for SDN Modelling}
\label{subsection:modelling}
An SDN is made of a controller linked with several devices, which are the switches. 
The controller are linked to the switches via a secure message channel which conveys message flows between the controller and the switches.
The switches are interconnected via a data channel which  conveys data packets.
Consequently, the elementary abstractions are the basic sets that represent:
the switches (SW\_ID),
the packets (PACKET),
the messages (MESG),
the packet headers (HEADER),
the states of a switch (SW\_STATE). 
The messages  have types and may contain packets: $mesgType \in{}  MESG \pfun{} MESGTYPE$, 
        $mesgPk \in{} MESG \pfun{}  PACKET$.

A packet has several headers (MAC source address, MAC destination address, MAC type, IP source address, IP destination address, IP protocol, transport source port, transport destination port), for simplification we consider only one of such header: $pHeader_i \in{} PACKET \pfun HEADER$. In the model these headers are specified like the function $pHeader_i$.
All the previous sets and constants are gathered in a \textsf{CONTEXT}, seen my a \textsf{MACHINE} which contains the variables and the previous typing predicate and properties respecrively in a \textsf{VARIABLES} and \textsf{INVARIANT}   clauses.
%

The SDN is a set of components that work concurrently in an asynchronous manner; we build a first global abstract model that simulates the functioning of this asynchronous system. The global abstract model will be the basis for the development of the components. \\

To structure this abstract model, we consider the data model and the discretisation of the behaviour (a set of observed events) of each of its components as a family of events. This is important for mastering the interaction between components and the forthcoming decomposition of the model.

\paragraph{Switches.}
Each switch has a flow table which contains the elementary behaviour of the switch according to the packets entering the switch. The behaviour of a switch is as follows: when it receives a message from the controller, it analyses the information inside the message and accordingly performs the instructions of the controller, for example updating its table, delivering a packet to a given port indicated in the message,  dropping a packet or buffering a packet contained in the message. When a switch receives a packet from another switch, either it forwards the packet to another switch according to the rules in its current table, or it forwards the packet to the controller if there is no rule matching the packet headers. 
Accordingly, we have a set of switches: $switches \subseteq{}  SW\_ID$.
Each switch has:
\begin{itemize}
\item a flow table  which may be empty or made of several entries: \\
    	$flowTable \in{} ENTRY \pfun{} switches$.\\
        Each entry has several headers (similar as for packets); each one is specified as follows:\\
        $eHeader_i \in{} ENTRY \pfun{}  HEADER$\\
        $dom(eHeader_i) = dom(flowTable)$
        
\item a status:  $swStatus \in{} SW\_ID \pfun{} SW\_STATE \land \dom(swStatus) = switches$
\item a buffer $swIncomingMsg$ containing all messages received by the switches:
$swIncomingMsg \subseteq{}  MESG \times switches$
\item a buffer $swIPk$ for all packet received by a switch, before treatment:\\
$swIPk \in PACKET \rel switches$;
$swIncomingPk$ is the set of packets such that
$swIncomingPk \subseteq{} PACKET$ and 
$swIncomingPk = \dom(swIPk)$.\\
Each packet has a header: $pHeader_i \in{} PACKET \pfun{}  HEADER$
\item a buffer $swOMsg$ that contains messages to be sent to the controller:\\
$swOMsg \in MESG \rel switches$;
$swOutgoingMsg$ is a set of messages such that
$swOutgoingMsg \subseteq{} MESG \land swOutgoingMsg = \dom(swOMsg)$

\item a buffer $swOPk$ containing all packets to be sent to other switches or to the controller:
$swOPk \in PACKET \rel switches$ and $swOutgoingPk$ the set of packets such that
$swOutgoingPk \subseteq PACKET \land swOutgoingPk = \dom(swOPk)$.
\end{itemize}

\paragraph{Behaviour of the switch.}
We capture the behaviour of the switch by considering how it is involved in the interaction with its environment. Each impact of the environment is considered as an event. The (re)actions of the switch are modelled as events that in turn impact or not the environment. We have then a set of events characterizing the switches; they are as follows.
\begin{description}
\item[\textsf{sw\_rcv\_machingPkt}:] the condition for the occurrence of this event is that there is in the incoming packets of a switch $sw$, a packet $pkt$, received from another switch via the data channel ($(pkt \mapsto sw) \in{} dataChan$), which header ($ahd = pHeader1(pkt)$) is matched with one entry of the flow table of $sw$:\\
($\exists ee . ( ((ee \in{} ENTRY) \land (ee \in \dom(flowTable)) ) ~\land~(eHeader1(ee) = ahd))$) ; 
the effect of the event is that the packet should be forwarded to another switch: $swIPk :=  swIPk \cup \{pkt\}$
\item[\textsf{sw\_rcv\_unmachingPkt}:] its occurs when a switch receives a packet (from another switch) which header does not match any entry of the flow table.
\item[\textsf{sw\_sndPk2ctrl}:] occurs when a switch emits a message containing an unmached packet to the controller;
\item[\textsf{sw\_sendPckt2sw}:]  a switch sends a packet to another switch via the data channel;
\item[\textsf{sw\_newFTentry}:] the occurrence of this event expresses that a new entry is added to the flow table.
\item[$\cdots$]
\end{description}

\paragraph{Controller.}
A controller is the device that administrates the switches using control messages.
It has buffers which contain messages or packets to be sent/received to/from switches:
a buffer for incoming packets ($ctlIncomingPk \subseteq{}  PACKET$);
a buffer for outgoing packets ($ctlOutgoingPk \subseteq{} PACKET$).

The controller  emits/receives messages on/from the secure channel. These messages contain either data packets or instructions to control the switches.
Among the control message we have:
the \textsf{Add} order to add an new entry into the table flow of a switch;
\textsf{Modf} to modify  an  entry into the table flow of a switch;
\textsf{Del} to delete an entry into the table flow of a switch.


\paragraph{Behaviour of the controller.} As for the switch, the behaviour of the controller is captured and modelled as a set of events denoting how the controller interacts with its environment. 
Each impact of the environment is considered as an event; the (re)actions of the controller are modelled as events that in turn impact or not the environment. 

As illustration, among the events of the controller we have the following:
\begin{description}
\item[\textsf{ctl\_emitPkt}:]  this event occurs when the controller emits to a switch $sw$, through a message, one of its pending packets; 
the condition for this occurrence is that there is some pending packets in the dedicated buffer ($pkt \in{} ctlOutgoingPk$).\\
The effect of the event is that a message containing the packet is added to the secure channel:
$secureChan:= secureChan \cup \{msg \mapsto sw\}$ and the buffer is updated: $ctlOutgoingPk := ctlOutgoingPk \backslash \{pkt\}$. Figure \ref{fig:ctlEvent} gives the Event-B specification of the event; all the remaining events are specified in a similar way.
\item[\textsf{ctl\_rcvPacketIn}:] 
 this event occurs when the controller receives a packet from a switch which previously received it but does not have an entry matching it.
\item[\textsf{ctl\_askBarrier}:] the occurrence of this event specifies when the controller asks for a barrier; that means the controller orders the switch to perform some control with urgency and to send a barrier acknowledgement.\\
\end{description}

\vspace{-0.4cm}
\begin{figure}[htb]
\begin{center}
\begin{boxedminipage}{12.5cm}
\vspace{-0.3cm}
\begin{tabbing}
\hspace{0.6cm}\=\hspace{1.0cm}\=\hspace{1.4cm}\=\hspace{1cm}\=\hspace{1cm}\\
\bkey{\textsf{event}}  \textsf{ctl\_emitPkt} // the controller emits a mesg conveying a packet\\
\bkey{\textsc{any}}  sw  pkt msg \\
\bkey{\textsc{where}} ~~~~~~~~  \bcmt{/* the  guard  */}\\
\>	 $sw \in{} switches$ // in destination to one of the switches\\
\>	 $pkt \in{} PACKET$  \\
\>	 $pkt  \in{} ctlOutgoingPk$ // one of the packet to be sent on the sw \\
\>   $msg \in{} MESG$ // a given message to convey the packet\\
\>	 $(msg \mapsto{} PKOut) \in{} mesgType$ // a packet of type OUT \\
\>	 $(msg \mapsto{} pkt) \in{} mesgPk$ // the message contains the packet\\
\bkey{\textsc{then}} ~~~~~~~~  \bcmt{/* the substitution */} \\
\> $secureChan := secureChan \cup \{msg \mapsto{} sw\}$ //emission on the channel\\
\> $ctlOutgoingPk := ctlOutgoingPk \setminus{} \{pkt\}$ \\
\bkey{\textsc{end}}
\end{tabbing}
\end{boxedminipage}
\end{center}
\vspace{-0.3cm}
\caption{Event-B specification of the event \textsf{ctl\_emitPkt}}
\label{fig:ctlEvent}
\end{figure}

The global abstract model comprises in an \textsf{EVENT} clause, all the  events characterizing the switches and the controller; the occurrence of each event is due to some conditions of the SDN and this occurrence has effect on the SDN. In Event-B a \textit{guard} captures each condition; an Event-B \textit{substitution} describes the effect of the event.

\paragraph{Interaction between Controller and Switches.}
The interaction is based on communications via channels; we distinguish a data packet channel and a control message channel.
The channels are modelled with sets.
A switch or a controller writes/reads messages on/from the channels according to their behaviour.

\noindent
\begin{boxedminipage}{12.2cm}
\begin{center}
        $secureChan \subseteq{} MESG \times switches$

        $dataChan \subseteq{} PACKET \times switches$
        \end{center}
\end{boxedminipage}

A first abstract Event-B model is obtained by gathering all these abstractions on data and behaviour.

\subsection{Model Construction Strategy: the Refinements}
Despite the general development strategy in Event-B which consists to build an abstract global model of a system and then to use several refinements to make it precise, it is still challenging to determine the refinement steps according to the problem at hand. In this work we have considered as one of our targets, the main components of the SDN. That is, we tried to deal with details related to the targeted components (switches, controller). The questions are: what characterizes the switches and what are the impacts on their environments? what characterizes the controller behaviour and what are the impacts on its environments? By answering these questions we went out by introducing, for instance, that switches use various ports and they receive/emit messages on ports. Consequently the first abstract model of channels is impacted and then refined.

We focus on the architecture of the SDN, and then tried to list the details that will support actually the achievement of the network services. We have listed, the detailed structure of packets, the structure of messages, the fine-grained processing of packets inside the switches. Then we order these details and tried to handle them one by one. It follows that we have to detail in the refinements:
the structure of packets with various headers and body parts;
the structure of messages, and accordingly the refinement of the abstract channels;
the behaviour of the events that specify the behaviour of both switches and controller.
This guided us to master the gradual modelling. From the methodological point of view this is a recipe for Event-B adepts.

We also follow the basic recommendations of Event-B to consider small steps of refinement at time.  
Table \ref{table:refinements} gives an overview of the refinement chains.
\noindent
\begin{table}
\begin{tabular}{|c|p{9.9cm}|}
\hline
\texttt{GblModel0} &  ~The first abstract model; all the events are specified at a high level; for instance we do not have yet information on ports, etc \\
 \hline
~\texttt{GblModel0\_1}~   &  ~Refinement. Ports and headers are introduced in the state space thus refined; the related events are refined.\\
 \hline
\texttt{GblModel0\_2}  &  ~Refinement. Priorities are introduced in the state space; messages are sent from the controller with a priority in their header.\\
 \hline
\texttt{GblModel0\_3} & Refinement. The events guard are refined according to priority rules\\
\hline
\end{tabular}

\caption{The refinement steps }
\label{table:refinements}
\end{table}

\subsection{Data Refinement}

The set of ports (PORTID\_ACTION) are introduced as data refinement details in the GblModel0\_1 refined  abstract model. Packets are sent on ports according to the actions defined in the entries of the flow table. One port (also called action) may be the destination of a set of packets. An entry may specify several actions or ports. 
The various fields in SDN packets are also introduced as data refinement with the functions: $macSrc$, $macDst$, $IpSrc$, $IpDst$, $IpProto$, $TpSrc$, $TpDst$, $TpSrcPt$, $TpDstPt$.

\noindent
\begin{boxedminipage}{12.2cm}
\begin{center}
        $actionsQueues \in{} ACTION \pfun{} \pow{}(PACKET)$ // packets targeting a port

        $actions \in{} ENTRY \pfun{} \pow{}(ACTION)$ // ports concerned by an entry 

        $dom(actions) = dom(flowTable)$ // all entries have target ports 

\end{center}
\end{boxedminipage}

\subsection{Behavioural Refinement}

\subsubsection{Explicit priority}
The controller (via a human administrator) can introduce \textit{priorities} as an information contained in the messages. Priorities are comparable, they are numbers. Consequently, we introduced this refinement level where the messages are refined by adding to them a field which represents their priority.
In Event-B, this is a function giving the priority of each message: $msgPriority \in{} MESG \pfun{} MSG\_PRIORITY$ where $MSG\_PRIORITY$ is the set of priorities (a subset of naturals).
Accordingly, the event \textsf{ctl\_emitPkt} for instance, is now refined in the model (\texttt{GblModel0\_2}); its substitution sets the priority of the message which is sent.

\subsubsection{Implicit priority}
We introduced implicit priorities via a partial order on messages to be sent; in the sequel the symbol $\prec$ denotes this partial order. 

To avoid inconsistencies in the behaviour of switches, the messages they sent should be reordered. In practice, when for instance the flow table is modified by an instruction coming from the controller, the outgoing packet in a switch may be forwarded in a wrong destination due to the modification. Besides, the controller can use the barrier to impose a quick modification.

Accordingly the modification messages coming from the switch should have less priority compared with the the forwarding messages. A priority rule which reorder the events, is that: the  \textsf{add} control messages are processed after the forwarding of all data packets. The involved events in the model are: \textsf{sw\_newFTentry}, \textsf{sw\_sendPckt2sw}, \textsf{sw\_sendPckt2Ctrl}. Therefore we have the following ordering:\\

\noindent
\begin{boxedminipage}{12.2cm}
\begin{center}
\textsf{sw\_newFTentry} $\prec$ \textsf{sw\_sendPckt2sw}

\textsf{sw\_sndPk2ctrl} $\prec$ \textsf{sw\_sendPckt2sw}

\textsf{sw\_sndPk2ctrl} $\prec$ \textsf{sw\_newFTentry}
\end{center}
\end{boxedminipage}

As far as the \textsf{Del} order is concerned, as lost packets in the network can be claimed, we use this hypothesis to consider that the \textsf{Del} order has priority on the forward packet.
For the \textsf{Add} order, this does not present an inconsistency risk for outgoing packets. For this reason the \textsf{Add} order can be processed in any order.
Barrier messages coming from the controller are the most priority ones.
Unmatched packets to be returned to the controller are less priority than the packet to be forwarded to other switches: a rule is that packets to the controller are sent if there is no packet to be forwarded to other switches.

These priorities have been implemented (in \texttt{GblModel0\_3}) as a refinement of our model. The guards of the involved events have been strengthen with this rules.


%% file: eventB_nutshell-1.tex
Event-B \cite{EventB_Abrial2010,DBLP:journals/scp/HoangKBA09} is a modelling and development method where components are modelled as abstract machines which are composed and refined into concrete machines. 
An \textit{abstract machine}  describes a mathematical model of a system behaviour\footnote{A system behaviour is a discrete transition system}.
In an Event-B modelling process, abstract machines constitute the dynamic part whereas \textit{Contexts} are used to describe the static part.  
 A \textit{Context} is seen by machines. It is made of carrier sets and constants. It may contain properties (defined on the sets and constants), axioms and theorems.
A machine is described, using named clauses, by a state space made of typed variables and invariants, together with several \textit{event} descriptions. 

\vspace{-0.2cm}
\paragraph{State Space of a Machine}
The variables constrained by the invariant (typing predicate, properties) describe the  state space of a machine.
The change from one state to the other is due to the effect of the events of the machine. Specific properties required by the model may be included in the invariant. The predicate $I(x)$ denotes the invariant of machine, with $x$ the state variables.

\vspace{-0.2cm}
 \paragraph{Events of an Abstract Machine}
Within Event-B, an event is the description of a system transition. Events are  spontaneous and show the way a system evolves. 
An event $e$ is modelled as a \textit{guarded substitution}: $e \defs eG \Longrightarrow eB$ where $eG$ is the event \textit{guard} and $eB$ is the event \textit{body} or \textit{action}.
An event may occur only when its guard holds. 
The action of an event describes, with simultaneous generalised substitutions, how the system state evolves when this event occurs: disjoint state variables are updated simultaneously.

The effect of events are modelled with generalised logical substitution (S) using the global variables and constants. For instance a basic substitution  \texttt{x := e}  is logically equivalent to the predicate \textit{x' such that x' = e}. This is symbolically written $ x' : (x' = e)$ where $x'$ corresponds to the state variable $x$ after the substitution and $e$ is an expression. In the rest of the paper, the variable $x$ is generalised to the list of state variables.

Several events may have their guards held simultaneously; in this case, only one of them  occurs. The system makes internally a nondeterministic choice. If no guard is true the abstract system is blocking (deadlock).

In Event-B \textit{Proof Obligations} are defined to establish model consistency via invariant preservation. Specific properties (included in the invariant) of a system are also proved in the same way.

\vspace{-0.2cm}
\paragraph{Refinement.} An important feature of the Event-B method is the availability of refinement technique to design concrete system from its abstract model by stepwise enrichment of the abstract model. During the refinement process new variables ($y$) are introduced; the invariant is strengthened without breaking the abstract invariant, and finally the events guards are strengthened. In the invariant $J(x,y)$ of the refinement, abstract variables ($x$) and concrete variables ($y$) are linked. The refinement is accompanied with proof obligations in order to prove their correctness with respect to the abstract model.

\vspace{-0.2cm}
\paragraph{\texttt{Rodin} Tool.} \texttt{Rodin}\footnote{{\small http://wiki.event-b.org/index.php/Main\_Page}}  is an open tool dedicated to building and reasoning on B models, using mainly provers and the \textsf{ProB} model-checker. \texttt{Rodin} is made of several modules (plug-ins) to work with B models and interact with related tools.

%% file: controller.tex
%
%
The purpose is to derive SDN components from the global model resulting from the chain of refinements; such derivation is enabled with Event-B via the  use of \textit{model decomposition} techniques: the Abrial'style decomposition (called the A-style
decomposition)~\cite{DBLP:journals/fuin/AbrialH07} based on shared
variables, and the Butler'style decomposition (called the B-style
decomposition)~\cite{DBLP:conf/ifm/Butler09,DBLP:conf/fmco/SilvaB10}
based on shared events.
In the A-style decomposition, which we have used, events are
first partitioned between Event-B sub-components and then, shared variables 
of these sub-components, but only modified by a sub-component, are used to introduce some \textit{external events} in the sub-components which do not modify the variables.
These external events simulate  the behaviour of the events which modify the variables, in the components where the variables are not modified.
To avoid inconsistency,  external events should not be refined.
In the B-style decomposition, variables are first partitioned
between the sub-components and then shared events (which use the
variables of both sub-components) are split between the sub-components
according to the used variables.
We used the A-style decomposition because it is more relevant when we
consider that the events describe the behaviour of each specific component (controller, switch) of the SDN.



\subsection*{Decomposition into a controller and  switches}

According to our modelling approach (see Section \ref{subsection:modelling}) where events are gathered by family, 
it is straightforward to list the events that describe the behaviour of the controller in order to separate them from the events related to the switches.
The controller component is made of all the events, already introduced as such and prefixed with \textsf{ctrl}, which simulate the behaviour of the controller (see Section \ref{subsection:modelling}): \textsf{ctl\_emitPkt},  \textsf{ctl\_rcvPacketIn},   \textsf{ctl\_askBarrier}, etc.   
Formally, the decomposition is as if a model ${\mathcal M_{\Sigma}}$ composed of components ${\mathcal S_{\sigma_1}}$ and ${\mathcal C_{\sigma_2}}$, such that ${\mathcal M_{\Sigma}} \vDash P$, is split into submodels ${\mathcal C_{\sigma_1}}$  and ${\mathcal S_{\sigma_{2}}}$ such that ${\mathcal C_{\sigma_1}} \vDash P$ and  ${\mathcal S_{\sigma_{2}}} \vDash P$, with $\Sigma = {\sigma_1} \cup {\sigma_2}$ the alphabet of  ${\mathcal M_{\Sigma}}$ and $\sigma_1$ (resp. $sigma_2$) the alphabet of ${\mathcal C_{\sigma_1}}$ (resp. ${\mathcal S_{\sigma_2}}$). 

We experimented with the decomposition plugin of the Rodin toolkit using the A-Style decomposition approach. 

A challenging issue here is the question of partitioning a set of identical behaviours;  
 for instance if we  would like to decompose the behaviours of the switches as a partition. This question is out of the scope of the existing decomposition techniques because of the non-determinism of data and event modelling. 

%% file: experimentation.tex
%
%
The global abstract model has been incrementally worked out by combining invariant verification, refinements and simulation.
This is done with the \textsf{Rodin} platform\footnote{http://wiki.event-b.org/index.php}. In Table \ref{table:statistics} we give the statistics (on proof obligations) of the performed proofs on the abstract model and its refinements; they were mostly automatically discharged by the \textsf{Rodin} prover. The complete model is available at \url{http://pagesperso.ls2n.fr/~attiogbe-c/mespages/nabla/sdn/SDN-WP2.pdf}

\begin{wraptable}[10]{r}{.62 \linewidth}
  \vspace{-0.6cm}
    \begin{tabular}{|l|l|l|l|l|}
      \hline
      Elt Name & Total & Auto & Manual & Undischarged \\
     \hline
     \hline
     \texttt{SDN-WP2} & 210 & 202 & 8 & 0\\
     \texttt{GlModel0} & 97 & 94 & 3 & 0 \\
     \texttt{GlModel0$\_1$} & 63 & 59 & 4 & 0 \\
     \texttt{GlModel0$\_{1\_1}$} & 2 & 2 & 0 & 0\\
     \texttt{GblModel$0\_2$} &  2 & 2 & 0 & 0\\
     \texttt{GblModel$0\_3$} &  0 & 0 & 0 & 0\\
      \hline
    \end{tabular}
  \caption{Proof statistics}
 \label{table:statistics}
\end{wraptable}

The model of the last refinement level has been decomposed into a controller component and switches which preserve all the proved properties.
One benefit of deriving components from a global formal model is the ability to study required properties involving the components and their environment.
We have illustrated this study with a few properties expected from SDN. Both safety and liveness properties have been considered.
This can be extended to other specific properties, following a similar approach.

We have expressed and proved several global properties on the model before its decomposition into components. For instance:
\textit{The data packets received by any switch are sent by the controller or by the other switches}.
\vspace{-0.2cm}
\paragraph{Proof.} Assume $ctlSentPkts$ be the set of packets sent by the controller; we have to prove that $\forall sw . (sw : switches \implies swIPs[\{sw\}] \subseteq ctlSentPks)$.
If $swIncomingPks$ is the union of the buffers of the switches, then it suffices to establish that $swIncomingPks \subseteq  ctlSentPkts$. \hfill $\qed$\\

Several such safety properties (e.g. Table \ref{table:safetyProp}) have been proved on the model.
\vspace{-0.2cm}
\begin{table}
\begin{center}
  \begin{tabular}{|l|l|}
    \hline
\textbf{\textsf{SP$_{a}$}} & ~Any packet in the data channel was sent by the controller or the switches\\
  \hline
\textbf{\textsf{SP$_{b}$}} & ~Any packet in the switches buffers was sent by the controller or the switches\\
  \hline
\textbf{\textsf{SP$_{c}$}} & ~The packets sent via the message channel  are contained in  $ctl\_sentPkts$\\
\hline
\end{tabular}
   \end{center}
  \caption{Proof statistics}
 \label{table:safetyProp}
\end{table}

\vspace{-0.2cm}
Similarly, liveness properties study is undertaken using stepwise checking of basic properties. For instance, we prove that, the data packets generated by the controller, are finally emitted by this later. The formula \textbf{ \textsf{LP$_{deliv}$}}  (see Table \ref{table:liveProp}) expresses this property. Literally it describes that after the occurrence of the event \texttt{ctl\_havePacket} we will fatally (\textsf{F}) observe the occurrence of \texttt{ctl\_emitPkt}. The other formula in Table \ref{table:liveProp} are similar; the \textsf{X} symbol  stands for the next operator.
Event-B provides, via the \textsf{ProB} tool integrated in the \textsf{Rodin},
the facilities to state and prove liveness properties.  \textsf{ProB} supports LTL, its extension LTL[e] and CTL properties with the standard modal and temporal operators.

\begin{table}
  \begin{center}
  \begin{tabular}{|l|>{\centering\arraybackslash}p{10cm}|}
    \hline
\textbf{ \textsf{LP$_{OKstatus}$} } & \textsf{e(ctl\_askStatusMsg) $\implies$ F(e(ctl\_rcvStatus))}\\
    \hline
\textbf{ \textsf{LP$_{deliv}$} } & \textsf{e(ctl\_havePacket) $\implies$ F(e(ctl\_emitPkt))}\\
    \hline
\textbf{ \textsf{LP$_{OKMach}$} }& \textsf{e(ctl\_emitPkt) $\implies$ X(e(sw\_rcv\_machingPkt))}\\
\hline
  \end{tabular}
    \end{center}
  \caption{A few liveness properties in LTL/ProB}
  \label{table:liveProp}
    \end{table}

%% file: conclusion.tex
%
%

We have shown how to build  correct controller and switches components from the refinement of a global formal model of an SDN system, using the decomposition of the global model  into the target components. The global model was first built by a systematic construction using refinements. The construction of the abstract model itself  was achieved so as to be reusable as a recipe for Event-B developers, following the steps we had identified. 
We overcame the challenging  modelling in Event-B, of an SDN system, viewing it as a discrete events system, and thus as an interaction between its main components.
We evaluated our model and components for their conformance to the properties required for SDN systems. We experimented the various aspect related to property proving and simulation, usig the \texttt{Rodin} tool. 
As far as we know, among the related work using formal approaches, this study is the first one proposing an event-based appoach for studying SDN systems. 

We provided a core event-based model to found frameworks dedicated to the development, analysis and simulation of SDN-based applications.

As future work, instead of a one-shot derivation of a specific code for the contoller,  we are investigating a parametric environment to enable the construction of specific controllers targeting various languages. The same idea is relevant for the switches. In light-weight distributed applications requiering the deployment of adhoc SDN, it is desirable to build various specific SDN switches from a single abstract model. Consequently a process of generic refinement to codes, will be beneficial.